\documentclass[a4paper,12pt,noffotinbib]{article}
\usepackage[english]{babel} 
\usepackage[T1]{fontenc}
\usepackage{float}
\usepackage{setspace}
\usepackage{appendix}
\usepackage{parcolumns}
\usepackage{cite}
\usepackage{times}
\usepackage[OT1]{fontenc}
\usepackage{type1cm}
\usepackage{url}
\usepackage{subfigure}
\usepackage{braket}
\usepackage{graphicx}
\usepackage{sidecap}
\usepackage{xspace}
\usepackage{tikz}
\usepackage{indentfirst}
\usepackage[mathscr]{eucal}
\usepackage{geometry}
\usepackage{booktabs}
\usepackage{indentfirst}
\usepackage{bm}
\usepackage{multirow}
\usepackage{dcolumn}
\usepackage{graphicx}
\usepackage{mathrsfs}  
\usepackage{enumerate}
\usepackage[utf8]{inputenc}
\usepackage{amsmath}
\usepackage{amssymb}
\usepackage{amsfonts}
\usepackage{amsthm}
\usepackage{mathrsfs}
\usepackage{graphicx}
\usepackage{multirow}
\usepackage{ulem}	
\usepackage{amsmath,mathtools}
\usepackage{comment}
\usepackage{amsbsy}
\usepackage{hyperref}
\usepackage{bookmark}
\makeatletter
\@addtoreset{equation}{section}

\makeatother
\newcommand{\tr}{{\rm tr}}

\newcommand{\w}[1]{\\[0.#1cm]}

\def\Tr{{\rm Tr}}
\def\tr{{\rm tr}}

\newcommand{\be}{\begin{equation}}
\newcommand{\ee}{\end{equation}}
\newcommand{\bea}{\begin{eqnarray}}
\newcommand{\eea}{\end{eqnarray}}
\newcommand{\vs}[1]{\vspace{#1 mm}}

\newcommand{\lc}{{\mit\Gamma}}
\newcommand{\con}{K}

\def\cD{{\cal D}}

\def\cL{{\cal L}}

\def\cO{{\cal O}}

\newcommand{\trT}[2]{\mathrm{tr}_{(#1#2)}T}
\newcommand{\divT}[1]{\mathrm{div}_{(#1)}T}

\newcommand{\trdivT}[1]{\mathrm{tr}\mathrm{div}_{(#1)}T}
\newcommand{\DivT}[1]{\mathrm{Div}_{(#1)}T}

\newcommand{\trDivT}[1]{\mathrm{tr}\mathrm{Div}_{(#1)}T}
\begin{document}
%\maketitle

\vs{3}
\begin{center}
{\large\bf On the renormalization of Poincar\'e gauge theories}
\vs{8}

{\large
Oleg Melichev${}^*$\footnote{e-mail address: omeliche@sissa.it}\ 
and Roberto Percacci${}^*$\footnote{e-mail address: percacci@sissa.it}\ 

\vs{8}
}

${}^*${International School for Advanced Studies, via Bonomea 265, I-34136 Trieste, Italy}
{and INFN, Sezione di Trieste, Italy}

\vs{5}
%%%%%%%%%%%%%%%%%%%%%%%%%%%%%%%
{\bf Abstract}
\end{center}
{\narrower\small
Poincar\'e Gauge Theories are a class of Metric-Affine Gravity 
theories with a metric-compatible (i.e. Lorentz) connection and with an action 
quadratic in curvature and torsion.
We perform an explicit one-loop calculation starting with a single term of each type
and show that not only are all other terms generated, but also many others.
In our particular model all terms containing torsion are redundant
and can be eliminated by field redefinitions, but there remains a new term
quadratic in curvature, making the model non-renormalizable.
We discuss the likely behavior of more general theories of this type.
}

%%%%%%%%%%%%%%%%%%%%%%%
\section{Introduction}
%%%%%%%%%%%%%%%%%%%%%%%

Metric-Affine theories of Gravity (MAGs) are a vast class of theories of gravity that
generalize Einstein's General Relativity (GR) by allowing the connection to be independent of the metric \cite{Utiyama:1956sy,Hehl:1976kj,Hehl:1994ue,Blagojevic:2013xpa}.
In the most general case, the connection will have both torsion and nonmetricity,
and the difference between the independent connection and the Levi-Civita connection is a tensor called distortion.
MAGs can be presented in various equivalent ways.
As in all theories of gravity, one can use either coordinate bases,
or orthonormal bases (a.k.a. tetrads, or vierbeins),
or even completely general frames.
This amounts to a choice of $GL(4)$ gauge
\cite{perbook1,shumen,komar,Dabrowski:1986cf,Neeman:1987jgg,higgsgr,siegel,kirsch,leclerc}.
Furthermore, one can choose the independent variables to be
either metric/frame and connection, in which case we say that the theory
is in its Cartan form,
or metric/frame and distortion, in which case we say that the theory
is in its Einstein form \cite{Baldazzi:2021kaf}.
\footnote{A MAG in Einstein form is basically a metric theory of gravity coupled to
some form of matter.}
These are mere field redefinitions, and leave the physics unchanged.
Thus a given MAG can be presented in (at least) six different equivalent ways.
Which of these forms is more convenient depends on what one wants to do,
and we shall see that the calculation of quantum effects can be made simpler
by choosing the appropriate formulation.

In a general MAG, the number of possible terms in the Lagrangian is huge:
there are 12 terms of dimension two
and close to one thousand terms of dimension four \cite{Baldazzi:2021kaf}.
It is clear that in order to study their properties,
one has to start by considering simpler subcases.
One can either impose additional symmetries, such as projective invariance
\cite{Afonso:2017bxr,Aoki:2019rvi,Percacci:2020ddy},
or kinematical constraints.
If the distortion is symmetric in the first and third index,
or antisymmetric in the second and third, we call the theory a
symmetric/antisymmetric MAG, respectively.
\footnote{If the distortion has both properties, it is zero.}
Symmetric MAGs have zero torsion and antisymmetric MAGs
have zero nonmetricity (i.e. the connection is metric-compatible).
As an example of further kinematical constraints, 
one can recover Weyl's gauge theory \cite{weyl,Scholz:2017pfo}
by imposing that either nonmetricity \cite{Ghilencea:2018dqd,Ghilencea:2020piz}
or torsion \cite{Lucat:2016eze,Iosifidis:2018zjj} be purely vectorial.
%In this paper we focus on antisymmetric MAGs,
%but some of the lessons learned will apply more generally.

A popular subclass of antisymmetric MAGs arises as follows.
One can view the frame field as a gauge field for translations,
and combine it with the Lorentz connection to form a Poincar\'e connection.
The natural Lagrangian for such a theory is quadratic in the curvature
of the Poincar\'e connection, i.e. quadratic in the curvature of the Lorentz
connection and quadratic in torsion
\cite{Hayashi:1973mj,Cho:1976fr,Hehl:1978yt,Hayashi:1979wj,Pilch:1979bi}.
Irrespective of the geometric motivation,
we shall call Poincar\'e gauge theories (PGTs) the subclass of antisymmetric MAGs
that have a Lagrangian quadratic in torsion and in the Lorentz connection,
possibly containing also a Palatini term (linear in curvature).

Little is known about the quantum properties of these theories,
that will depend on the choice of Lagrangian.
Their renormalizability has been discussed to some extent
in \cite{Tseytlin:1981nu,Lin:2018awc,Lin:2019ugq}.
We recall that in the metric theories containing squares of curvature,
commonly known as Quadratic Gravity (QGs), a naive perturbative analysis
yields either renormalizable but non-unitary theories (containing the square
of the Riemann or Ricci tensor) or unitary but non-renormalizable theories
(quadratic in the Ricci scalar).
\footnote{Here unitarity is taken to mean absence of ghosts,
but there are arguments to the effect that
the presence of ghosts in these theories does not spoil unitarity
\cite{Mannheim:2006rd,Anselmi:2018ibi,Anselmi:2018tmf,Donoghue:2019fcb,Salvio:2014soa}.}
One would expect that in general the same will be true also for MAGs.
Classes of ghost- and tachyon-free MAGs have been found in \cite{neville,sezgin,sezgin2,Percacci:2020ddy}, and are of measure zero.
%Thus it seems likely that the question of renormalizability of MAGs
%will not have an entirely straightforward answer.
To the best of our knowledge, there are very few explicit statements about
one loop divergences and beta functions in PGT (or, more generally, any MAG).
Among them we count: \cite{Tseytlin:1981nu},
that used the results of \cite{Deser:1974xq}
in a MAG containing terms of the form $R+F^2$
(where $F$ is the curvature of the independent connection),
\cite{Floreanini:1991cw}, that used a kind of mean field approximation,
\cite{Percacci:2011uf}, that was limited to the calculation of effective potentials,
and \cite{Donoghue:2016vck}, where the fluctuations of the metric are neglected
and diffeomorphism invariance is not fixed.
In the present paper we consider the simplest PGT,
with only one term quadratic in curvature and one quadratic in torsion.
We calculate its logarithmic divergences and find that
not only are all other terms of the PGT generated,
but essentially all dimension-four terms
and, somewhat surprisingly, also other higher-dimensional powers of curvature.
Thus the theory is certainly non-renormalizable off shell.
We then discuss the role of field redefinitions and find that many
of the new terms can be eliminated in this way, but not all.
The conclusion is that the model we started with is non-renormalizable
also on shell.
The calculation suggests various ways in which the conclusions
could change for different choices of action.

There are also some lessons to be learned at the technical level.
The calculation of quantum effects in these theories is very cumbersome,
because the fields have many components.
One can reduce the number of fields by working in coordinate frames
but this is a relatively small advantage:
the 16 components of the tetrad are reduced to the 10 components of
the metric, but the number of independent components of the connection
remains 24.
On the other hand, we find that in order to reduce the effective action
to the trace of the logarithm of a minimal Laplace-type operator,
we have to work with tetrad variables
and fix the Lorentz gauge in a manner that is close to Yang-Mills theories.
Furthermore, even though we start with the theory in its Cartan form,
it is convenient to work with the Levi-Civita connection.
Thus the logarithmically divergent terms are first found in Einstein form,
and then converted back to Cartan form.

In Section 2 we present general facts about the kinematics and dynamics of PGTs.
In Section 3 we set up the one loop calculation.
Intermediate steps of the calculations are reported in Appendices,
but the final results are discussed in Section 4.
The effect of field redefinitions is discussed in Section 5,
and Section 6 contains a summary of possible extensions.

Preliminary results of this paper have appeared in O. M.'s Ph.D. thesis
\cite{Melichev:2023lpn}.

%%%%%%%%%%%%%%%%
\section{Poincar\'e gauge theories}
%%%%%%%%%%%%%%%%

In this section we collect well known facts about antisymmetric MAGs
and in particular Poincar\'e gauge theories.
This is meant mainly to establish our notation and terminology.
In the end of this section we specify the action that we are going
to take as a starting point for our quantum calculation.

\subsection{Kinematics}

We use orthonormal frames $\{e_a\}$ in the tangent spaces
and $\{e^a\}$ in the cotangent spaces.
They are related to the coordinate bases by
$e_a=\theta_a{}^\mu\partial_\mu$ and 
$e^a=\theta^a{}_\mu dx^\mu$.
The components of the metric in the orthonormal frames
are $\eta_{ab}\equiv \mathrm{diag}(-1,1,1,1)$ and the components of a Lorentz connection are $A_\lambda{}^a{}_b$.
They are related to the components in the coordinate bases by
\begin{eqnarray}
\label{pullbackmetric}
g_{\mu\nu}&=&\theta^a{}_\mu\, \theta^b{}_\nu\, \eta_{ab}\\
\label{pullbackconnection}
A_\lambda{}^\mu{}_\nu&=&\theta_a{}^\mu A_\lambda{}^a{}_b \theta^b{}_\nu
+\theta_a{}^\mu \partial_\lambda \theta^a{}_\nu
\end{eqnarray}
The components $A_{abc}=\theta_a{}^\lambda\eta_{bd}A_\lambda{}^d{}_c$ 
are antisymmetric in $b,c$.

The curvature is defined by the commutator of covariant derivatives:
\be
[D_v,D_w]X-D_{[v,w]}X=F(v,w)X\ ,
\label{defF}
\ee
where $D_v X^a=v^\mu(\partial_\mu X^a+A_\mu{}^a{}_c X^c)$
and $[v,w]$ is the Lie bracket.
The definition is such that the derivatives of $v$ and $w$ cancel and
$F(v,w)X^a=v^\mu w^\nu F_{\mu\nu}{}^a{}_c X^c$, with
\be
F_{\mu\nu}{}^a{}_b=
\partial_\mu A_\nu{}^a{}_b
-\partial_\nu A_\mu{}^a{}_b
+A_\mu{}^a{}_c A_\nu{}^c{}_b
-A_\nu{}^a{}_c A_\mu{}^c{}_b\ .
\ee

While $D_v X$ is a tensor of the same type as $X$,
one can define also the covariant differential $DX$,
which is a tensor with one more covariant index,
and $DDX$, that has two more covariant indices.
In components, it is customary to write $(DDX)^a_{\mu\nu}=D_\mu D_\nu X^a$.
The antisymmetric part of this tensor is
\be
2(DDX)^a_{[\mu\nu]}
=[D_\mu,D_\nu]X^a=F_{\mu\nu}{}^a{}_c X^c
-T_\mu{}^\lambda{}_\nu D_\lambda X^a\  ,
\ee
which is different from (\ref{defF}).
Thus, one has to be careful about the meaning of ``commutator of covariant derivatives''.

The tetrad uniquely defines the Levi-Civita connection,
whose coordinate components $\lc_\mu{}^\rho{}_\sigma$
are the Christoffel symbols, and are related to the tetrad components
$\lc_\mu{}^a{}_b$ as in (\ref{pullbackconnection}).
For the LC connection, the covariant derivative is denoted $\nabla$ and
the curvature $R$ defined by
\be
[\nabla_v,\nabla_w]X-\nabla_{[v,w]}X=R(v,w)X\ ,
\label{defR}
\ee
is the Riemann tensor.
Finally, the last ingredient is the torsion tensor, 
which is defined as the exterior covariant derivative of the frame field:
\be
\label{torsion}
T_\mu{}^a{}_\nu=
\partial_\mu \theta^a{}_\nu-\partial_\nu \theta^a{}_\mu+
A_\mu{}^a{}_b\, \theta^b{}_\nu-A_\nu{}^a{}_b\, 
\theta^b{}_\mu
\ee
and in coordinate frames is just
\be
T_\mu{}^\rho{}_\nu=
A_\mu{}^\rho{}_\nu-A_\nu{}^\rho{}_\mu\  .
\ee

Given a metric $g_{\mu\nu}$, 
a Lorentz connection can be uniquely decomposed into
\be
A_{\alpha\beta\gamma}=\lc_{\alpha\beta\gamma}+K_{\alpha\beta\gamma}
\label{changevar}
\ee
where $\lc_{\alpha\beta\gamma}$ (in a coordinate basis) are the Christoffel symbols
and $K_{\alpha\beta\gamma}$,
antisymmetric in $(\beta,\gamma)$, is called the contorsion tensor.
It is related to the torsion by
\bea
\label{Tphi}
T_{\alpha\beta\gamma} &=& K_{\alpha\beta\gamma}-K_{\gamma\beta\alpha}\ ;
\\
\con_{\alpha\beta\gamma} & = & \frac{1}{2}\left(T_{\alpha\beta\gamma}+T_{\beta\alpha\gamma}-T_{\alpha\gamma\beta}\right)\ .
\eea
The curvature tensor of the Levi-Civita connection is the Riemann tensor 
$R_{\mu\nu}{}^\alpha{}_\beta$ and is related to the curvature tensor
of the independent connection as follows:
\bea
F_{\mu\nu}{}^\alpha{}_\beta & = & 
R_{\mu\nu}{}^\alpha{}_\beta
+\nabla_{\mu}K_{\nu \,\,\, \beta}^{\,\,\, \alpha}
-\nabla_{\nu}K_{\mu\,\,\, \beta}^{\,\,\, \alpha}
+K_{\mu\,\,\, \gamma}^{\,\,\, \alpha}K_{\nu \,\,\, \beta}^{\,\,\, \gamma}
-K_{\nu \,\,\, \gamma}^{\,\,\, \alpha}K_{\mu \,\,\, \beta}^{\,\,\, \gamma}\ .
%\nonumber\\
%&=&R_{\mu\nu}{}^\alpha{}_\beta
%+\nabla_{\mu}T_{\nu \,\,\, \beta}^{\,\,\, \alpha}.....
\label{Fphi}
\eea
The analog of the Ricci scalar for the connection $A_\mu{}^\alpha{}_\beta$
is the unique contraction $F=F_{\mu\nu}{}^{\mu\nu}$,
which, up to total derivatives, can be written as
\bea
F & = & R
%+\nabla_{m}K_{n}^{\,\,\,\, mn}-\nabla_{n}K_{m}^{\,\,\,\, mn}
+K_{\mu \,\,\,\, \gamma}^{\,\,\,\, \mu}K_{\nu}^{\,\,\,\, \gamma\nu}
-K_{\nu\mu\gamma}K^{\mu\gamma\nu}\ .
\\
& = & R
+\frac{1}{4}T_{\alpha\beta\gamma}T^{\alpha\beta\gamma}
 +\frac{1}{2}T_{\alpha\beta\gamma}T^{\alpha\gamma\beta}
-T_{\alpha}{}^{\alpha\beta}T^{\; \gamma}{}_{\gamma\beta}\ .
\label{palatiniTQ}
\eea

One can define four different notions of ``total covariant derivative'' of the tetrad,
using either $A$ or $\lc$ for the latin and the greek index.
We observe that the transformation (\ref{pullbackconnection})
is equivalent to the statement that the total covariant derivative of the tetrad,
using the same connection on both indices, is zero
(this is often called the ``tetrad postulate'').
The total covariant derivative, 
using the Levi-Civita connection
for the greek index and the dynamical connection for the latin one,
is nothing but the contorsion:
\be
\cD_\mu\theta^a{}_\rho\equiv
\partial_\mu\theta^a{}_\rho-\lc_\mu{}^\sigma{}_\rho\theta^a{}_\sigma
+A_\mu{}^a{}_b\theta^b{}_\rho=K_\mu{}^a{}_\rho\ .
\label{totcovd}
\ee

The following table summarizes our notation,
which is Yang-Mills-like for the independent connection,
and GR-like for the Levi-Civita connection,
and we stress that the same notation is used
when the connection coefficients refer to a coordinate basis (greek indices)
or an orthonormal basis (latin indices):
\begin{center}
\begin{tabular}{|c|c|c|c|}
\hline
& coefficients 
& cov. der.
& curvature 
\\
\hline
Levi-Civita connection 
&  $\lc_\mu{}^\rho{}_\sigma$
& $\nabla_\mu$ 
& $R_{\mu\nu}{}^\rho{}_\sigma$
\\
\hline
Independent connection 
&  $A_\mu{}^\rho{}_\sigma$
& $D_\mu$ 
& $F_{\mu\nu}{}^\rho{}_\sigma$
\\
\hline
\end{tabular}
\end{center}

%%%%%%%%%%%%%%%%%%%%%%%%%%%%%%%
\subsection{Dynamics}
%%%%%%%%%%%%%%%%%%%%%%%%%%%%%%%

The standard Lagrangian for PGT is quadratic in curvature and torsion:
\bea
\cL\!\!\!\!&=&\!\!\!\!
-\frac12\Big[ -a_0 F
+ T^{\mu\rho\nu} \big(a_1 T_{\mu\rho\nu} 
+ a_2 T_{\mu\nu\rho}\big)  
+ a_3 T^\mu T_\mu
\nonumber\w2
&&  
+ F^{\mu\nu\rho\sigma} \big( c_1 F_{\mu\nu\rho\sigma} 
+ c_3 F_{\rho\sigma\mu\nu} 
+ c_4 F_{\mu\rho\nu\sigma} \big)
+F^{\mu\nu} \big(c_7 F_{\mu\nu} + c_8 F_{\nu\mu} \big)
+c_{16}F^2
\Big]\ ,
\label{action}
\eea
where $T_\mu= T_\lambda{}^\lambda{}_\mu$.
The non-consecutive numbering of coefficients comes from \cite{Baldazzi:2021kaf}
and is motivated by compatibility with the action of more general MAGs.
In writing this action, we have made two choices.
One can write the action in orthonormal bases, simply changing the
middle index of torsion and the last two indices of curvature from greek to latin.
This is a mere gauge choice and is completely inconsequential
to the physical content of the theory.
We have chosen to think of the action as a functional of the metric
and of the independent gauge field $A$.
This is the choice of variables that makes MAG more similar to a YM theory,
and in \cite{Baldazzi:2021kaf} we called it the Cartan form of the theory.
One can choose to present the theory in what we called the Einstein form,
where the action is regarded as a functional of the metric and
torsion (or equivalently metric and contorsion).
This change of dynamical variables is performed by using (\ref{changevar}).
\footnote{
Note that whereas in Cartan form torsion is a derived quantity,
being constructed from the connection (in coordinate frames) 
or from the connection and tetrad (in orthonormal frames),
in Einstein form it has to be regarded as an independent field.}

In the following we shall consider the case where the only nonzero couplings
are $c_1$ and $a_1$.
The Lagrangian in orthonormal frames can then be written
\begin{equation}
\mathcal{L}=-\tfrac{1}{2} \sqrt{|g|} g^{\mu\rho}g^{\nu\sigma}\eta_{ab}
\left(c_1 F_{\mu \nu}{}^a{}_c F_{\rho\sigma}{}^b{}_d \eta^{cd} 
%+ a_4 Q_{\rho \mu \nu } Q^{\rho \mu \nu } 
+ a_1 T_\mu{}^a{}_\nu T_\rho{}^b{}_\sigma \right)
\label{LagMin}
\end{equation}
This writing clearly exposes the variables that have to be varied.
Here the metric is to be regarded as a composite of the tetrads,
as in (\ref{pullbackmetric}) and $F$ depends on $A$ but not on the tetrad.

%Applying the technique of the spin projectors one finds that the
%theory has an accidental gauge symmetry in the spin/parity 1$^-$ sector,
%and propagates various modes with spin/parity 2$^+$,
%2$^-$, 1$^+$, 1$^-$, 0$^+$, 0$^-$. (CHECK...)

This action looks much more complicated when written in Einstein form:
\begin{equation}
\cL^E=-\tfrac{1}{2} \sqrt{|g|} 
\left(\mathcal{L}^E_2+\mathcal{L}^E_3+\mathcal{L}^E_4\right)\ .
\label{LagMinE}
\end{equation}
In the basis defined in \cite{Baldazzi:2021kaf}, we have
\bea
\mathcal{L}^E_2&=&c_1 \Big[R_{\mu\nu\rho\sigma}R^{\mu\nu\rho\sigma}
+5R^{\mu\nu}\nabla^\rho T_{\rho\mu\nu}
+\frac32 \nabla_\alpha  T_{\mu\nu\rho}\nabla^\alpha T^{\mu\nu\rho}
+ \nabla_\alpha  T_{\mu\nu\rho}\nabla^\alpha T^{\mu\rho\nu}
\\
&&
\!\!\!\!
+ \nabla^\alpha  T_{\alpha\mu\nu}\nabla_\beta (-T^{\beta\mu\nu}+T^{\beta\nu\mu})
-\frac12 \nabla^\alpha  T_{\mu\alpha\nu}\nabla_\beta T^{\mu\beta\nu}
-2 \nabla^\alpha  T_{\alpha\mu\nu}\nabla_\beta T^{\mu\beta\nu}\Big]
+a_1 T_{\mu\nu\rho}T^{\mu\nu\rho}\ .
\nonumber
\eea
while the Lagrangian $\mathcal{L}_3$ contains several terms of the form
$RTT$ and $TT\nabla T$,
and the Lagrangian $\mathcal{L}_4$ contains several terms of the form
$TTTT$, all of which we do not write out explicitly.
The terms in $\mathcal{L}_2^E$ affect the propagation in flat space,
while those in $\mathcal{L}_3^E$ and $\mathcal{L}_4^E$ only give rise
to 3- and 4-point vertices.

Even though the Lagrangian is presented in Cartan form,
it is easier to work with Levi-Civita covariant derivatives $\nabla$
rather than the $A$-covariant derivatives $D$.
As for tensorial quantities such as
curvatures $F$ and $R$, they can coexist in formulas.
To read off the divergences and beta functions, we will first bring everything
to Einstein form, and finally convert everything back to Cartan form,
in a particular basis of invariants.

%%%%%%%%%%%
\section{Setup for one loop calculation}
%%%%%%%%%%%

The calculations have been performed in the Euclidean regime,
where the action differs from (\ref{LagMin}) by an overall sign.

%%%%%%%%%%%
\subsection{Expansion}
%%%%%%%%%%%

The basic variables in (\ref{LagMin}) are the tetrad and connection.
Their variations will be called
\be
\delta\theta^a{}_\mu=X^a{}_\mu\ ,\qquad
\delta A_\mu{}^a{}_b=Z_\mu{}^a{}_b\ .
\ee
Then we have
\bea
\delta \sqrt{g}&=&\sqrt{g}X^\rho{}_\rho
\\
\delta^2 \sqrt{g}&=&\sqrt{g}\left(X^\rho{}_\rho X^\sigma{}_\sigma
-X^{\rho\sigma}X_{\sigma\rho}\right)
\\
\delta g^{\mu\nu}&=&-X^{\mu\nu}-X^{\nu\mu}
\\
\delta^2g^{\mu\nu}&=&2(X^{\mu\rho}X^\nu{}_\rho
+X^{\mu\rho}X_\rho{}^\nu+X^{\nu\rho}X_\rho{}^\mu)
\\
\delta F_{\mu \nu}{}^a{}_b&=&
\bar D_\mu Z_\nu{}^a{}_b-\bar D_\nu Z_\mu{}^a{}_b
+\bar T_\mu{}^\rho{}_\nu Z_\rho{}^a{}_b
\\
\delta^2 F_{\mu \nu}{}^a{}_b&=&
2[Z_\mu,Z_\nu]^a{}_b
\\
\delta T_\mu{}^a{}_\nu&=&
\bar D_\mu X^a{}_\nu-\bar D_\nu X^a{}_\mu
+Z_\mu{}^a{}_\nu-Z_\nu{}^a{}_\mu
+\bar T_\mu{}^\rho{}_\nu X^a{}_\rho
\\
\delta^2 T_\mu{}^a{}_\nu&=&
2(Z_\mu{}^a{}_b X^b{}_\nu-Z_\nu{}^a{}_b X^b{}_\mu)\ .
\eea
Varying the action and using these relations one arrives at the Hessian,
which is a quadratic form in $X$ and $Z$.
It is convenient to rewrite all $D$ derivatives as $\nabla$ derivatives
plus terms linear in torsion.
Furthermore, since the Hessian is a combination of tensors and their
covariant derivatives, we can write it in coordinate bases,
by converting all latin indices to greek ones.
The terms with two derivatives are
\be
a_1 X_{\mu\nu}(-\bar\nabla^2 g^{\nu\sigma}+\bar\nabla^\sigma\bar\nabla^\nu)X^\mu{}_{\sigma}
+c_1 Z_\mu{}^{\rho\sigma}(-\bar\nabla^2 g^{\mu\nu}+\bar\nabla^\nu\bar\nabla^\mu)Z_{\nu\rho\sigma}\  ,
\label{nonmin}
\ee
where the bar over the covariant derivatives indicate that they
are computed with the background metric.
The occurrence of the nonminimal terms would greatly complicate the calculation,
but can be avoided by choosing a suitable gauge.
We observe that this is only possible because we started in the vierbein formalism
and the Hessian vanishes on fields that are local Lorentz or diffeomorphism transformations.
Starting in coordinate bases, one would have to fix only 
the diffeomorphism invariance
(4 parameters) and one can fix this gauge in such a way as to remove the
nonminimal terms in the $X$-$X$ sector.
By choosing a suitable Lorentz gauge fixing we can
remove also the nonminimal terms in the $Z$-$Z$ sector.
We shall see this in detail in Section 3.3.

%%%%%%%%%%%
\subsection{Gauge algebra}
%%%%%%%%%%%

Due to the structure of the gauge group,
the gauge fixing conditions for gravity in tetrad formulation
(whether the connection is independent or not) is more complicated
than imposing separate gauge conditions for diffeomorphisms and local Lorentz tranformations.
This kind of complication already occurs in the case of Yang-Mills fields
coupled to gravity \cite{Jackiw:1978ar,Daum:2009dn}.
In the case of Einstein-Cartan theory it has been discussed in 
\cite{Daum:2010qt,Benedetti:2011nd,Daum:2010zz,Daum:2013fu}.
We will broadly follow these references.

The fields of antisymmetric MAG are defined on $OM$, 
the bundle of orthonormal frames of the base manifold $M$,
and its associated bundles, and the action is invariant under automorphisms of this bundle.
One can parametrize locally this group by diffeomorphisms of $M$ and 
local Lorentz transformations,
acting in the standard way on latin and greek indices, respectively:
\bea
\delta^L_\omega\theta^a{}_\mu&=&-\omega^a{}_b\theta^b{}_\mu\ ,
\nonumber\\
\delta^L_\omega A_\mu{}^a{}_b&=&D_\mu\omega^b{}_c\ ,
\nonumber\\
\delta^D_v\theta^a{}_\mu&=&\cL_v \theta^a{}_\mu
=v^\rho\partial_\rho\theta^a{}_\mu+\theta^a{}_\rho\partial_\mu v^\rho\ ,
\nonumber\\
\delta^D_v A_\mu{}^a{}_b&=&\cL_v A_\mu{}^a{}_b\ ,
\eea
where $\omega_{ab}=-\omega_{ba}$ is an infinitesimal Lorentz gauge parameter
and $v^\mu$ is an infinitesimal diffeomorphism (a vectorfield on $M$).
Note that the latin indices are inert under this definition of diffeomorphism.
The algebra of these transformations is
\bea
\left[\delta^L_{\omega_1},\delta^L_{\omega_2}\right]&=&\delta^L_{[\omega_1,\omega_2]}
\nonumber\\
\left[\delta^D_{v_1},\delta^D_{v_2}\right]&=&-\delta^D_{[v_1,v_2]}
\nonumber\\
\left[\delta^D_{v},\delta^L_{\omega}\right]&=&\delta^L_{\cL_v\omega}
\eea
This shows that the local Lorentz transformations are a normal
subgroup of the full gauge group, and the diffeomorphisms are the quotient
of the full group by this subgroup.

Now we observe that whereas the general fluctuation $X^a{}_\mu$ transforms properly
under local Lorentz transformations, the gauge fluctuation $\delta_v\theta^a{}_\mu$ 
does not.
This would become a serious obstacle in the following, 
because $\delta_v\theta^a{}_\mu$ is used in the
construction of the ghost operator, and this definition would lead to a
non-covariant ghost operator.
\footnote{
One can also try to covariantize $\cL_v \theta^a{}_\mu$ by adding and subtracting
$v^\rho\lc_\rho{}^\nu{}_\mu\theta^a{}_\mu$.
However, the resulting covariant derivatives are only covariant under diffeomorphisms:
the derivative acting on $\theta^a{}_\mu$ is not Lorentz-covariant,
with the result that $\cL_v \theta^a{}_\mu$ is not a Lorentz vector.}
The solution consists in defining a modified action of diffeomorphisms
on the fields, which consists of the original action defined above,
plus an infinitesimal Lorentz transformation with parameter
$\epsilon^a{}_b=-v^\mu A_\mu{}^a{}_b\equiv -(v\cdot A)^a{}_b$.
\footnote{Here we follow \cite{Daum:2010qt,Daum:2010zz,Daum:2013fu}.
Alternatively one could also use 
$\epsilon^a{}_b=-v^\mu \lc_\mu{}^a{}_b\equiv -(v\cdot A)^a{}_b$,
where $\lc_\mu{}^a{}_b$ are the components of the Levi-Civita connection
in the orthonormal frame.
This would lead to a simpler transformation for $\theta$ 
(the second term would be absent)
but a more complicated one for $A$.}
\be
\tilde \delta^D_v=\delta^D_v-\delta^L_{v\cdot A}\ .
\label{modified}
\ee
The action of these modified diffeomorphisms on the fields is
\bea
\tilde\delta^D_v\theta^a{}_\mu&=&
\theta^a{}_\rho\nabla_\mu v^\rho
+v^\rho K_\rho{}^a{}_\mu
\ ,
\nonumber\\
\tilde\delta^D_v A_\mu{}^a{}_b&=&v^\rho F_{\rho\mu}{}^a{}_b\ ,
\eea
where we used (\ref{totcovd}).
Their algebra is
\bea
%\left[\delta_{\omega_1},\delta_{\omega_2}\right]&=&-\delta_{[\omega_1,\omega_2]}
%\nonumber\\
\left[\tilde\delta^D_{v_1},\tilde\delta^D_{v_2}\right]&=&
-\tilde\delta^D_{[v_1,v_2]}-\delta^L_{F(v_1,v_2)}
\nonumber\\
\left[\tilde\delta^D_{v},\delta_{\omega}\right]&=&0\ ,
\label{modalg}
\eea
where $F(v_1,v_2)^a{}_b=v_1^\mu v_2^\nu F_{\mu\nu}{}^a{}_b$.
This is just a different way of parametrizing the full gauge group
of the theory, where the normal subgroup has remained untouched.

In background field calculations we have to define how to split the
transformation of a field into transformations of its background and fluctuation parts.
In the so called ``quantum'' transformations $\delta^Q$ the backgrounds are invariant
and the whole transformation of the field is attributed to the fluctuation:
\bea
\delta^{QL}_\omega \bar\theta^a{}_\mu&=&0\ ,
\nonumber\\
\delta^{QL}_\omega \bar A_\mu{}^a{}_b&=&0\ ,
\nonumber\\
\delta^{QL}_\omega X^a{}_\mu&=&-\omega^a{}_b\,\theta^b{}_\mu\ ,
\nonumber\\
\delta^{QL}_\omega Z_\mu{}^a{}_b&=&D_\mu\omega^a{}_b\ ,
\eea
\bea
\tilde\delta^{QD}_v\bar\theta^a{}_\mu&=&0\ ,
\nonumber\\
\tilde\delta^{QD}_v \bar A_\mu{}^a{}_b&=&0\ ,
\nonumber\\
\tilde\delta^{QD}_v X^a{}_\mu&=&
\theta^a{}_\rho\bar\nabla_\mu v^\rho+v^\rho K_\rho{}^a{}_\mu\ ,
\nonumber\\
\tilde\delta^{QD}_v Z_\mu{}^a{}_b&=&v^\rho F_{\rho\mu}{}^a{}_b\ ,
\eea
The ``background'' transformations $\delta^B$ are defined in such a way that
the backgrounds transform
as the original field (in particular, $\bar A$ transforms as a connection).
In detail, the background Lorentz transformations are
\bea
\delta^{BL}_\omega\bar\theta^a{}_\mu&=&
-\omega^a{}_b\,\bar\theta^b{}_\mu
\ ,
\nonumber\\
\delta^{BL}_\omega \bar A_\mu{}^a{}_b&=&\bar D_\mu\omega^a{}_b\ ,
\nonumber\\
\delta^{BL}_\omega X^a{}_\mu&=&
-\omega^a{}_c \,X^c{}_\rho
\ ,
\nonumber\\
\delta^{BL}_\omega Z_\mu{}^a{}_b&=&
-\omega^a{}_c\, Z_\mu{}^c{}_b
+Z_\mu{}^a{}_c \,\omega^c{}_b
\ ,
\eea
and the background diffeomorphisms are given by the Lie derivative on all fields.
The background diffeomorphisms can be covariantized as above, in particular
\bea
\bar\delta^{BD}_v\bar\theta^a{}_\mu&=&\bar\theta^a{}_\rho\bar\nabla_\mu v^\rho
+v^\rho \bar K_\rho{}^a{}_\mu\ ,
\nonumber\\
\bar\delta^{BD}_v \bar A_\mu{}^a{}_b&=&v^\rho \bar F_{\rho\mu}{}^a{}_b\ ,
%\nonumber\\
%\bar\delta^{BD}_v X^a{}_\mu&=&
%X^a{}_\rho\bar\nabla_\mu v^\rho\ ,
%\nonumber\\
%\bar\delta^{BD}_v Z_\mu{}^a{}_b&=&v^\rho\bar\nabla_\rho Z_\mu{}^a{}_b
%-Z_\rho{}^a{}_b \bar\nabla_\mu v^\rho\ ,
\eea

%%%%%%%%%%%
\subsection{Gauge fixed Hessian}
%%%%%%%%%%%

We fir the gauge by choosing the Lorentz-like gauge conditions
\bea
\chi_D^\mu&=&\bar\nabla^\nu X^\mu{}_\nu
\\
\chi_L{}^a{}_b&=&\bar\nabla^\nu Z_\nu{}^a{}_b\  .
\eea
In the latter expression it is understood that the covariant derivative
is defined in terms of the background Levi-Civita connection for both types of indices.
Then, the gauge fixing action is
\be
S_{GF}=\int d^4x\sqrt{|\bar g|}\left[
\frac{a_1}{\alpha_D}\bar g_{\rho\sigma}\chi_D^\rho\chi_D^\sigma
+\frac{c_1}{\alpha_L}\eta_{ac}\eta^{bd}
\chi_L{}^a{}_b \chi_L{}^c{}_d\right] \ .
\ee
This breaks invariance under the ``quantum'' transformations
while preserving invariance under the ``background''  transformations. 
Since the total background covariant derivative of the background tetrad is zero,
in the second term we can harmlessly transform all the latin indices to greek ones.
Then we remain only with background diffeomorphism invariance,
and we do not need to worry about the covariantization that was discussed
in the previous section.
That discussion only plays a role in the definition of the ghost action.

Setting the parameters $\alpha_D=\alpha_L=1$ (Feynman gauge),
integrating by parts and commuting derivatives one gets
\bea
S_{GF}&=&\int d^4x\sqrt{|\bar g|}\Big[
-a_1 X_{\mu\nu}\bar\nabla^\sigma\bar\nabla^\nu X^\mu{}_\sigma
-c_1 Z_\mu{}^{\rho\sigma}\bar\nabla^\nu\bar\nabla^\mu Z_{\nu\rho\sigma}
\label{gaugefixing}
\\
&&
+a_1 X_{\mu\nu}\bar R^{\nu\sigma}X^\mu{}_\sigma
-a_1 X_{\mu\nu}\bar R^{\mu\rho\nu\sigma}X_{\rho\sigma}
+c_1 Z_{\mu\alpha\beta}\bar R^{\mu\sigma}Z_\sigma{}^{\alpha\beta}
-2c_1 Z_{\mu\alpha\beta}\bar R^{\mu\rho\alpha\sigma}Z_{\rho\sigma}{}^{\beta}
%+Z_{\mu\alpha\beta}\bar R^{\mu\rho\beta\sigma}Z_{\rho\alpha\sigma}
\Big] \ .
\nonumber
\eea
We see that the first line exactly cancels the unwanted nonminimal terms
in the (\ref{nonmin}).

At this point, rescaling 
\be
X^a{}_\mu \to \frac{1}{\sqrt{a_1}}X^a{}_\mu\ ,\qquad
Z_\mu{}^a{}_b\to \frac{1}{\sqrt{c_1}}Z_\mu{}^a{}_b
\label{rescaling}
\ee
and performing some integrations by parts,
one can write the gauge-fixed Hessian in the form
%\be
%H=\frac12\int d^4x\sqrt{g}
%\begin{pmatrix}
%X & Z
%\end{pmatrix}
%\begin{pmatrix}
%-\nabla^2+V_{XX}^\mu\nabla_\mu+W_{XX} & V_{XZ}^\mu\nabla_\mu+W_{XZ}\\
%V_{ZX}^\mu\nabla_\mu+W_{ZX} & -\nabla^2+V_{ZZ}^\mu\nabla_\mu+W_{ZZ}
%\end{pmatrix}
%\begin{pmatrix}
%X\\Z
%\end{pmatrix}
%\ee
%where the $V$'s and $W$'s are matrices in the space of the fields,
%with the appropriate free indices.
%Note that this contains al the terms coming from the second variation of the action,
%plus the terms in the second line of (\ref{gaugefixing}).
%More compactly
\begin{equation}
H = \frac12(\Psi, \cO \Psi).
\end{equation}
where $\Psi=\begin{pmatrix}
X\\Z
\end{pmatrix}$ and
\begin{equation}
\label{KinopGeneral}
\cO = -{\nabla}^2 \mathbb{I}+ \mathbb{V}^\sigma \nabla_\sigma+ \mathbb{W},
\end{equation}
with
\be
 \mathbb{V}^\mu=
\begin{pmatrix}
V_{XX}^\mu & V_{XZ}^\mu\\
V_{ZX}^\mu & V_{ZZ}^\mu
\end{pmatrix}
\qquad\qquad
\mathbb{W}=
\begin{pmatrix}
W_{XX} & W_{XZ}\\
W_{ZX} & W_{ZZ}\ ,
\end{pmatrix}
\ee
where the $V$'s and $W$'s are matrices in the space of the fields,
with the appropriate free indices.
Note that this contains all the terms coming from the second variation of the action,
plus the terms in the second line of (\ref{gaugefixing}).

The operator $\cO$ must be self-adjoint, which implies the conditions
\bea
V_{XX}^{[\mu\nu]\lambda[\alpha\beta]}
&=&-V_{XX}^{[\alpha\beta]\lambda[\mu\nu]}
\\
V_{XZ}^{[\mu\nu]\lambda[\alpha\beta\gamma]}
&=&-V_{ZX}^{[\alpha\beta\gamma]\lambda[\mu\nu]}
\\
V_{ZZ}^{[\mu\nu\rho]\lambda[\alpha\beta\gamma]}
&=&-V_{XX}^{[\alpha\beta\gamma]\lambda[\mu\nu\rho]}
\\
W_{XX}^{[\mu\nu][\alpha\beta]}
&=&W_{XX}^{[\alpha\beta][\mu\nu]}
-\nabla_\lambda V_{XX}^{[\alpha\beta]\lambda[\mu\nu]}
\\
W_{XZ}^{[\mu\nu][\alpha\beta\gamma]}
&=&W_{ZX}^{[\alpha\beta\gamma][\mu\nu]}
-\nabla_\lambda V_{ZX}^{[\alpha\beta\gamma]\lambda[\mu\nu]}
\\
W_{ZZ}^{[\mu\nu\rho][\alpha\beta\gamma]}
&=&W_{ZZ}^{[\alpha\beta\gamma][\mu\nu\rho]}
-\nabla_\lambda V_{ZZ}^{[\alpha\beta\gamma]\lambda[\mu\nu\rho]}
\eea

In these formulae, the square brackets are only meant to highlight
the grouping of indices, and do not represent antisymmetrization.

With the rescaling (\ref{rescaling}) the quantum fields $X$ and $Z$ have canonical
dimension one, $\mathbb{V}$ has dimension 1 and $\mathbb{W}$ has dimension 2.
We do not give the components of these tensors but just indicate the
general structures that they contain:
\bea
V_{XX}&\sim& T
\nonumber
\\
V_{XZ}&\sim& V_{ZX}\sim \left(\sqrt{\frac{a_1}{c_1}}\ ,\ \sqrt{\frac{c_1}{a_1}}F\right)
\nonumber
\\
V_{ZZ}&\sim& T
\label{Vform}
\\
W_{XX}&\sim& \left(T^2\ ,\  \nabla T\ ,\ \frac{c_1}{a_1}F^2\right)
\nonumber
\\
W_{XZ}&\sim& W_{ZX}\sim \left(\sqrt{\frac{a_1}{c_1}}T\ ,\ \sqrt{\frac{c_1}{a_1}}T F\right)
\nonumber
\\
W_{ZZ}&\sim& \left(\frac{a_1}{c_1}\ ,\ F\ ,\ T^2\ ,\ \nabla T\right)\ .
\label{Wform}
\eea
Here terms without tensors have to be understood as combinations of the metric.
We note that $\mathbb{V}$ and $\mathbb{W}$ can be treated as small perturbations
compared to the leading $\bar\nabla^2\approx E^2$ term, provided the
couplings and the background fields satisfy
\be
\frac{a_1}{c_1}\ll E^2\ ,\quad
T\ll E\ ,\quad
\nabla T\ll E^2\ ,\quad
F\ll\sqrt{\frac{a_1}{c_1}}E\ll E^2\ ,\quad
\ee
These conditions are necessary for the applicability of the
small-time heat kernel expansion.

%%%%%%%%%%%
\subsection{Ghost action}
%%%%%%%%%%%

The gauge fixing has to be supplemented by the ghost action.
We define the ghost operators
\be
\delta^{QL}_\Sigma\chi_L=\Delta_{LL}\Sigma\  ,\quad
\tilde\delta^{QD}_C\chi_L=\Delta_{LD}C\  ,\quad
\delta^{QL}_\Sigma\chi_D=\Delta_{DL}\Sigma\  ,\quad
\tilde\delta^{QD}_C\chi_D=\Delta_{DD}C\  .
\label{ghostops}
\ee
Here we have the infinitesimal ``quantum'' transformations applied to
the gauge fixing conditions, with the transformation parameters
$\omega^a{}_b$ and $v^\mu$ replaced by the ghost fields $\Sigma^a{}_b$ and $C^\mu$.
Then the ghost action is given by
\bea
S_{gh}&=&\int d^4x\sqrt{\bar g}
\left[
\bar\Sigma(\delta^{QL}_\Sigma\chi_L+\tilde\delta^{QD}_C\chi_L)
+\bar C(\delta^{QL}_\Sigma\chi_D+\tilde\delta^{QD}_C\chi_D)
\right]
\nonumber\\
&=&
\int d^4x\sqrt{\bar g}
\begin{pmatrix}
\bar\Sigma& \bar C
\end{pmatrix}
\begin{pmatrix}
\Delta_{LL} & \Delta_{LD}\\
\Delta_{DL} & \Delta_{DD}
\end{pmatrix}
\begin{pmatrix}
\Sigma\\ C
\end{pmatrix}\ ,
\eea
where $\bar\Sigma_a{}^b$ and $\bar C_\mu$ are the antighost fields.
All indices here have been suppressed for notational clarity.
When one evaluates explicitly the infinitesimal transformations in (\ref{ghostops}),
one obtains some operators of the form $\bar\nabla\nabla$,
i.e. containing both the background and the full connection.
However, we are ultimately only interested in the effective action at zero
fluctuation fields, so we can identify the full and background fields.
This means that the ghost operators are constructed entirely with
background fields.
Since the total covariant derivative of the background vierbein
with the background LC connection is zero, we can write
all formulas using only coordinate (greek) indices, without producing new terms.
The ghost operators are then:
\bea
(\Delta_{LL}\Sigma)^\alpha{}_\beta&=&\bar\nabla^2\Sigma^\alpha{}_\beta
+\bar\nabla^\nu(\bar K_\nu{}^\alpha{}_\gamma\Sigma^\gamma{}_\beta
-\Sigma^\alpha{}_\gamma\bar K_\nu{}^\gamma{}_\beta)
\nonumber\\
(\Delta_{LD}C)^\alpha{}_\beta&=&
\bar\nabla^\nu(\bar F_{\rho\nu}{}^\alpha{}_\beta C^\rho)
\nonumber\\
(\Delta_{DL}\Sigma)^\alpha&=&-\bar\nabla^\nu\Sigma^\alpha{}_\nu
\nonumber\\
(\Delta_{DD}C)^\alpha&=&\bar\nabla^2C^\alpha
+\bar\nabla^\nu(K_\rho{}^\alpha{}_\nu C^\rho)\ .
\label{fpgh}
\eea

%%%%%%%%%%%%%%%%%%%%%%%%%%%%%%%%
\section{One loop divergences and beta functions}
%%%%%%%%%%%%%%%%%%%%%%%%%%%%%%%%

\subsection{One loop divergences}

The one loop effective action is given by the classical action plus a quantum contribution
\be
\Gamma=S+\Delta\Gamma^{(1)}\ .
\ee
The effective action could contain non-local terms,
but these are related to infrared effects.
We are interested here in the UV behavior of the theory
and in particular in the logarithmically divergent part,
which is local.
Thus we can write
\be
\Gamma=\sum_i g_i\int d^4x\sqrt{g}\,\cL_i \ ,
\ee
where $\cL_i$ are operators constructed with the fields
and their covariant derivatives, and $g_i$ are the corresponding
(renormalized) dimensionless couplings.
A basis for a class of dimension-four operators $\cL_i$ will be given in Section 4.2.
There is a similar expansion for the classical action $S$,
whose coefficients $g_{Bi}$ are the bare couplings.
In the presence of a momentum cutoff $\Lambda$,
the logarithmically divergent part of $\Gamma$ can be written
\be
\Delta\Gamma^{(1)}_{log}=-B\log\left(\frac{\Lambda}{\mu}\right)\ ,
\ee
where $\mu$ is a reference scale that has to be introduced for dimensional reasons, and
\be
B=\sum_i \int d^4x\sqrt{g}\,\beta_i\cL_i\ ,
\ee
where the coefficients $\beta_i$ are defined by this equation.
Thus
\be
g_i(\mu)=g_{Bi}(\Lambda)-\beta_i\log\left(\frac{\Lambda}{\mu}\right)\ .
\ee
Here we assume that the bare couplings depend on the UV cutoff in such a way
that the renormalized couplings are finite, and as a consequence
the renormalized couplings must depend on $\mu$.
Then we find that
\be
\mu\frac{\partial{g_i}}{\partial\mu}=\beta_i\ ,
\ee
relating the coefficients of the logarithmic divergences to the beta functions.
%Note that we can also think of $B$ as
%\be
%B=-\Lambda\frac{\partial\Delta\Gamma^{(1)}_{log}}{\partial\Lambda}\ .
%\ee
%and therefore is a part of the ``beta functional'' 
%$\Lambda\frac{\partial\Gamma}{\partial\Lambda}$.

For our theory,
\be
\Delta\Gamma^{(1)}=\frac12\Tr\log\cO-\Tr\log\Delta_{gh}\ .
\ee
where both $\cO$ and $\Delta_{gh}$ are operators
of the form 
\be
-\bar\nabla^2+\mathbb{V}^\mu\bar\nabla_\mu+\mathbb{W}\ .
\ee
Defining $\nabla'=\nabla+\frac12 \mathbb{V}$ and redefining $\mathbb{W}$,
the term with one derivative can be removed.
Note that the connection $\nabla'$ now has vectorial torsion and nonmetricity,
but we can still apply the standard formulae for the Seeley-De Witt
coefficients, which give the following logarithmically divergent part of $\Tr\log\cO$:
\bea
B_\cO%\!\!\!\!\!\!\!\!\!\!
\!\!\!&=&\!\!\!-\frac{1}{(4\pi)^2}\int d^4x\sqrt{g}\,\tr\Big\{
\tfrac{1}{180}\left(\bar R_{\mu\nu\rho\sigma}\bar R^{\mu\nu\rho\sigma}
-\bar R_{\mu\nu}\bar R^{\mu\nu}+\tfrac52 \bar R^2\right) \mathbb{I}
\nonumber\\
&&\qquad\quad\quad
+\tfrac12\mathbb{W}^2
-\tfrac12 \mathbb{W}\bar\nabla_\mu \mathbb{V}^\mu
+\tfrac14\, \mathbb{W} \mathbb{V}_\mu \mathbb{V}^\mu
%\nonumber\\
%&&\qquad\qquad\qquad
%+{\color{red}\tfrac{1}{12}\bar R_{\mu\nu} \mathbb{V}^\mu \mathbb{V}^\nu}
-\tfrac16\bar R\, \mathbb{W}
+\tfrac{1}{12}\bar R\,\bar\nabla_\mu \mathbb{V}^\mu
-\tfrac{1}{24}\bar R\, \mathbb{V}_\mu \mathbb{V}^\mu
\nonumber\\
&&\qquad\quad\quad
+\tfrac{1}{12}\Omega_{\mu\nu}\Omega^{\mu\nu}
-\tfrac16\,\Omega_{\mu\nu}\bar\nabla^\mu \mathbb{V}^\nu
+\tfrac{1}{24}\,\Omega_{\mu\nu}[\mathbb{V}^\mu, \mathbb{V}^\nu]
\label{logdiv}
\\
&&\qquad\quad\quad
+\tfrac{1}{8}\bar\nabla_\mu \mathbb{V}^\mu\bar\nabla_\rho \mathbb{V}^\rho
-\tfrac{1}{8}\bar\nabla_\mu \mathbb{V}^\mu \mathbb{V}_\rho \mathbb{V}^\rho
+\tfrac{1}{32} \mathbb{V}_\mu \mathbb{V}^\mu \mathbb{V}_\rho \mathbb{V}^\rho 
\nonumber\\
&&\qquad\quad
+\tfrac{1}{24}(\bar\nabla_\mu \mathbb{V}_\nu-\bar\nabla_\nu \mathbb{V}_\mu)\bar\nabla^\mu \mathbb{V}^\nu
-\tfrac{1}{24}\,\bar\nabla_\mu \mathbb{V}_\nu[\mathbb{V}^\mu,\mathbb{V}^\nu]
+\tfrac{1}{192}\,[\mathbb{V}_\mu,\mathbb{V}_\nu][\mathbb{V}^\mu,\mathbb{V}^\nu]
\Big\}\ .
\nonumber
\eea
This formula can also be obtained directly from the expansion
of $\Tr\log$ and application of the off-diagonal heat kernel techniques
explained in \cite{Groh:2011dw}.
This calculation will be described in a forthcoming paper \cite{meli2}.

The individual terms in these expressions have been evaluated
for the operators $\cO$ and $\Delta_{gh}$
using the packages {\tt xAct} and {\tt xTensor}.
Summing them leads to the separate $X$-$Z$ and ghost contributions
that are far too large to be reported here.
In appendices A.1 and A.2 we give only the terms proportional
to two powers of curvature and torsion, which contribute to the flat space propagator.
The sum of all terms of this type generated by the fields $X$-$Z$ and the ghosts are
given in equations (\ref{XZtotal}) and (\ref{ghosttotal}) respectively.
These will be sufficient to illustrate some of our main points.
Besides such terms, the logarithmic divergences contain
most if not all terms of dimension four, as well as terms
of higher dimension, as we shall see.

%%%%%%%%%
\subsection{Bases}
%%%%%%%%%

We will show explicitly only the logarithmically divergent terms that are
proportional to two powers of curvature and torsion (and their derivatives).
To exhibit them we shall make use of the bases of invariants
introduced in \cite{Baldazzi:2021kaf}.
In the Einstein form of the theory, up to integrations by parts,
one has the following invariants:

\be
H^{RR}_1=R_{\mu\nu\rho\sigma}R^{\mu\nu\rho\sigma}\ ,\quad
H^{RR}_2=R_{\mu\nu}R^{\mu\nu}\ ,\quad
H^{RR}_3=R^2\ ,\quad
\label{RR}
\ee
\be
\begin{tabular}{lll}
$H^{TT}_{1} = \nabla^\alpha T^{\beta\gamma\delta} \nabla_\alpha T_{\beta\gamma\delta}$ \ ,
& $H^{TT}_{2} = \nabla^\alpha T^{\beta\gamma\delta} \nabla_\alpha T_{\beta\delta\gamma}$ \ ,
\\
$H^{TT}_{3} = \nabla^\alpha\trT12^\beta \nabla_\alpha\trT12_\beta$ \ ,
\\
$H^{TT}_{4} = \divT1^{\alpha\beta}\divT1_{\alpha\beta}$ \ ,
& $H^{TT}_{5} = \divT1^{\alpha\beta}\divT1_{\beta\alpha}$ \ ,
\\
$H^{TT}_{6} = \divT2^{\alpha\beta}\divT2_{\alpha\beta}$ \ ,
& $H^{TT}_{7} = \divT1^{\alpha\beta}\divT2_{\alpha\beta}$ \ ,
\\
$H^{TT}_{8} = %- \divdivT12^\alpha\trT12_\alpha$
\divT2^{\alpha\beta}\nabla_\alpha\trT12_\beta$ \ ,
& $H^{TT}_{9} = (\trdivT1)^2$ \ ,
\label{nablaTnablaT}
\end{tabular}
\ee
\be
\begin{tabular}{lll}
$H^{RT}_{1} = R^{\alpha\beta\gamma\delta} 
\nabla_\alpha T_{\beta\gamma\delta}$ \ ,
& $H^{RT}_{2} = R^{\alpha\gamma\beta\delta} 
\nabla_\alpha T_{\beta\gamma\delta}$ \ ,
&\\
$H^{RT}_{3} = R^{\beta\gamma} \divT1_{\beta\gamma}$ \ ,
& $H^{RT}_{4} = R^{\alpha\beta}\nabla_\alpha\trT12_{\beta}$ \ ,
& $H^{RT}_{5} = R\,\trdivT1$ \ .
\label{RnablaT}
\end{tabular}
\ee
Here $\trT12^b=T_a{}^{ab}$, $\divT3^{\alpha\beta}=\nabla_\gamma T^{\alpha\beta\gamma}$ etc. 
The Bianchi identities give three linear relations among the last five invariants.
We choose $\{H^{RT}_3 \ , \ H^{RT}_5 \}$
as independent invariants of type $R\nabla T$.

In the Cartan form of the theory the list of invariants is longer:
\bea
L^{FF}_1\!\!&=\!\!&F^{\mu\nu\rho\sigma}F_{\mu\nu\rho\sigma}\ , \quad
L^{FF}_3=F^{\mu\nu\rho\sigma}F_{\rho\sigma\mu\nu} \ ,\quad
L^{FF}_4=F^{\mu\nu\rho\sigma}F_{\mu\rho\nu\sigma}\ ,\quad
\nonumber
\\
L^{FF}_7&=&F^{(13)\mu\nu}F^{(13)}_{\mu\nu}\ ,\quad
L^{FF}_8=F^{(13)\mu\nu}F^{(13)}_{\nu\mu}\ ,\quad
L^{FF}_{16}=F^2\ .
\label{HFF}
\eea
\be
\begin{tabular}{lll}
$L^{TT}_{1} = D^\alpha T^{\beta\gamma\delta} D_\alpha T_{\beta\gamma\delta}$ \ ,
& $L^{TT}_{2} = D^\alpha T^{\beta\gamma\delta} D_\alpha T_{\beta\delta\gamma}$ \ ,
\\
$L^{TT}_{3} = D^\alpha \trT12^\beta D_\alpha\trT12_\beta$ \ ,
\\
$L^{TT}_{4} = \DivT1^{\alpha\beta}\DivT1_{\alpha\beta}$ \ ,
& $L^{TT}_{5} = \DivT1^{\alpha\beta}\DivT1_{\beta\alpha}$ \ ,
\\
$L^{TT}_{6} = \DivT2^{\alpha\beta}\DivT2_{\alpha\beta}$ \ ,
& $L^{TT}_{7} = \DivT1^{\alpha\beta}\DivT2_{\alpha\beta}$ \ ,
\\
$L^{TT}_{8} = \DivT2^{\alpha\beta}D_\alpha\trT12_\beta$ \ ,
& $L^{TT}_{9} = (\trDivT1)^2$ \ .
\end{tabular}
\label{HTT}
\ee
\be
\begin{tabular}{lll}
&$L^{FT}_1=F^{\mu\nu\rho\sigma}D_\mu T_{\nu\rho\sigma}$\,,\quad
$L^{FT}_3=F^{\mu\nu\rho\sigma}D_\mu T_{\rho\nu\sigma}$\,,
\\
& $L^{FT}_4=F^{\mu\nu\rho\sigma}D_\rho T_{\mu\nu\sigma}$\,,\quad
$L^{FT}_5=F^{\mu\nu\rho\sigma}D_\rho T_{\mu\sigma\nu}$\,,
\\
&$L^{FT}_8=F^{(13)\mu\nu} D_\mu\trT12_{\nu}$ \ ,
$L^{FT}_9=F^{(13)\mu\nu} D_\nu\trT12_{\mu}$ \ ,
\\
& $L^{FT}_{13}=F^{(13)\mu\nu}\, \DivT1_{\mu\nu}$ \ ,\quad
$L^{FT}_{14}=F^{(13)\mu\nu}\, \DivT1_{\nu\mu}$ \ ,
\\
&$L^{FT}_{17}=F^{(13)\mu\nu}\, \DivT2_{\mu\nu}$ \ ,
 $L^{FT}_{21}=F\, \trDivT1$ \ .
\end{tabular}
\label{HFT}
\ee
The non-consecutive numbering is due to the fact that in \cite{Baldazzi:2021kaf}
we defined bases for more general (not necessarily antisymmetric) MAGs.
Compared to the Einstein form of the theory, there are more invariants,
but also more relations, so that the number of independent invariants is the same.

Since we are mainly interested in PGT, it is natural to keep in the basis
all terms that contain $F$, and instead remove others.
We can choose as a basis the six $L^{FF}$ invariants, 
that are present in (\ref{action}), plus 
\be\label{carantisymbasisFF}
\{L^{TT}_1 \ , \ L^{TT}_2 \ , \ L^{TT}_3\ , \ L^{TT}_5 \} \quad
\mathrm{and}
\quad \{ L^{FT}_1 \ , \ L^{FT}_{8} \ , \ L^{FT}_9 \ , \ L^{FT}_{13}   \} \ .
\ee
By listing these terms we see that restricting ourselves to the
PGT Lagrangian (\ref{action}), we ignore eight possible
dimension-four term that can modify the propagation of torsion
and introduce dynamic mixing between the spin-two components
of torsion and metric.

%%%%%%%%%
\subsection{Results}
%%%%%%%%%

Summing $B_{XZ}-2B_{gh}$ from (\ref{XZtotal}) and (\ref{ghosttotal}) 
and removing the linearly dependent operators,
we arrive at the following expression of the divergent terms in Einstein form
\bea
B_E&=&
-\frac12\frac{1}{(4\pi)^2}\int d^4x\sqrt{g}\Big[
\frac{239}{60}H^{RR}_1
+\frac{31}{10}H^{RR}_2
-\frac{25}{12}H^{RR}_3
+\frac{131}{6}H^{RT}_3
-\frac{9}{4}H^{RT}_5
\nonumber\\
&&\qquad\qquad\qquad\qquad
+\frac{7}{3}H^{TT}_1
+\frac{13}{6}H^{TT}_2
-\frac23 H^{TT}_3
-2 H^{TT}_4
+\frac{7}{6}H^{TT}_5
\nonumber\\
&&\qquad\qquad\qquad\qquad
-\frac{43}{24}H^{TT}_6
-\frac{53}{12}H^{TT}_7
+\frac{14}{3} H^{TT}_8
+\frac{7}{6}H^{TT}_9
+\ldots\Big]\ .
\label{LogDivE2}
\eea
The ellipses stand for higher powers of $R$, $T$ and their $\nabla$-derivatives.

Then going to the basis in Cartan form we arrive at
\bea
B_C&=&
-\frac12\frac{1}{(4\pi)^2}\int d^4x\sqrt{g}\Big[
\frac{71}{15}L^{FF}_1
+\frac{97}{60}L^{FF}_3
-\frac{71}{15}L^{FF}_4
%\nonumber\\
%&&\qquad\qquad\qquad\qquad\ \ \ 
+\frac{107}{30}L^{FF}_7
-\frac{7}{15}L^{FF}_8
-\frac{25}{12}L^{FF}_{16}
\nonumber\\
&&\qquad\qquad\qquad\qquad
-\frac{253}{30}L^{FT}_1
+4L^{FT}_8
-\frac{299}{30}L^{FT}_9
-\frac{39}{5}L^{FT}_{13}
\nonumber\\
&&\qquad\qquad\qquad\qquad
+\frac{19}{30}L^{TT}_1
-\frac{1}{5}L^{TT}_2
-\frac{7}{30}L^{TT}_3
+\frac{32}{15}L^{TT}_5
+\ldots\Big]\
\label{logdivC2}
\eea
The ellipses stand for higher powers of $F$, $T$ and their $D$-derivatives.
This is our main result.
It shows that starting from a single $FF$-type term,
we not only generate all other $FF$ terms,
but also all other terms of dimension four that can affect the flat space propagator,
namely terms of the form $DTDT$ and $FDT$.
This is a sign that PGT is non-renormalizable off shell.
We shall discuss shortly what happens on shell,
but let us mention here that the ellipses include most, if not all,
dimension four terms, which are of the form
$TTF$, $TTDT$, $TTTT$,
as well as higher-dimensonal invariants of the form $FFFF$.
To understand the origin of the latter,
we go back to equations (\ref{Vform},\ref{Wform}) and note that
$V$ contains terms of the form $F\sqrt{c_1/a_1}$
and $W$ contains terms of the form $F^2 c_1/a_1$.
Then the terms in (\ref{logdiv}) proportional to $V^4$ or $W^2$
will contain terms of the form $F^4(c_1/a_1)^2$.
In the perturbative treatment of the theory,
the coupling is proportional to the inverse of $c_1$,
so the appearance of these terms is somewhat surprising.

We can understand better their origin by comparing with the situation
in QG, with Lagrangian of the form (schematically)
$m_P^2 R+\alpha R^2$, where $m_P$ is th Planck mass and
$\alpha$ is the inverse of the higher derivative coupling.
This is just the Einstein form of our action (\ref{LagMinE}),
with $T^2$ replaced by the Palatini term and neglecting the torsion.
In order to have a canonically normalized graviton,
one has to rescale the metric fluctuation by $1/\sqrt{\alpha}$.
Then, the kinetic operator is of the form
$\Box^2+\mathbb{V}^{\mu\nu}\nabla_\mu\nabla_\nu+\mathbb{W}$,
where $\mathbb{V}$ contains terms proportional to $R$ or $m_P^2/\alpha$
and $\mathbb{W}$ contains terms proportional to $R^2$ or $R m_P^2/\alpha$
\cite{Falls:2020qhj}.
In both cases (PGT and QG) the non-derivative part of the kinetic operator contains terms quadratic in curvature, that originate (for example) from the variation
of the determinant of the metric.
In the case of QG these terms come without any couplings,
because the coupling outside the action has been absorbed by the
rescaling of the metric fluctuation.
In the case of our MAG, the coupling outside the action is $c_1$
but the graviton is rescaled by $\sqrt{a_1}$,
and a dimensionful prefactor remains.
In QG the operator is of fourth order and the logarithmic divergences
contain at most $\tr \mathbb{W}$, \cite{barth,leepacshin,Gusynin:1988zt}
so terms of order $F^4$ cannot be generated.
In the case of the specific MAG considered here,
the operator is of second order and the logarithmic divergences
contain $\tr\mathbb{W}^2$, so that $F^4$ terms can, and do, appear.

%%%%%%%%%%%%%%
\section{On shell divergences}
%%%%%%%%%%%%%%

Next, we have to consider the effect of going on shell,
or equivalently of performing field redefinitions to remove 
some of the divergences.
We recall that in GR the equation of motion implies the vanishing
of the Ricci tensor. This allows us to discard the divergences
proportional to $R_{\mu\nu}R^{\mu\nu}$ and $R^2$.
Similarly, in our MAG the lower order equations of motion imply that $T=0$.
All divergences that involve $T$ vanish on shell,
which means that all terms involving $T$ are redundant.

To see this from another angle, consider any term of the form $T_{abc}X^{abc}$
where $X^{abc}$ is any combination of $T$, $\nabla T$, $F$ and $\nabla F$,
antisymmetric in $a$, $c$.
Varying the connection we have
$$
\delta_A T_\mu{}^a{}_\nu=\delta A_\mu{}^a{}_b\theta^b{}_\nu
-\delta A_\nu{}^a{}_b\theta^b{}_\mu\ .
$$
Simply choosing $\delta A_\mu{}^a{}_b=\tfrac12\theta^c{}_\mu X_c{}^a{}_b$,
we have $\delta_A T=X$.
Thus, applying this field redefinition of $A$ to the term $T^2$ in the action,
we can cancel any divergence that contains $T$,
confirming that those terms are redundant.

Thus, in principle, the only important logarithmic divergences are those
that involve curvatures,
and since torsion vanishes we can consider just $R$-curvatures
(as opposed to $F$-curvatures, that can form many more invariants).
Thus the set of possible counterterms in this theory is the same as
in a generic metric theory of gravity.
Looking at (\ref{LogDivE2}), we see that all three invariants
quadratic in $R$-curvatures are generated.
The Riemann-squared term can be absorbed by renormalizing the coupling $c_1$.
Another term can be removed using the Gauss-Bonnet theorem.
Unlike in GR, however, the equations of motion do not imply
the vanishing of the Ricci tensor,
and therefore one divergent term remains, implying that the theory is
on shell non-renormalizable already at one loop.

\section{More general theories}

From these results one can infer the likely behavior of other MAGs.
Let us start from other PGTs with action (\ref{action}),
but without the Palatini term (i.e. $a_0=0$).
Since the structure of the kinetic operator (\ref{KinopGeneral}),
with (\ref{Vform},\ref{Wform}), is generic,
one would expect that starting from ``almost any'' PGT
one will generate most dimension-four terms
as well as terms up to $F^4 c_i/a_j$.
It will still be the case that field redefinitions can eliminate
divergences containing torsion,
but now, given that generically all three $RR$ terms will be present in the bare Lagrangian,
all divergences of type $RR$ can be absorbed in redefinitions 
of the respective couplings.
The on-shell non-renormalizability of the PGT will therefore be due
to the higher powers of $F$, that do not seem likely to be
removable by redefining the metric.

Introducing in MAG the Palatini term will generate new contributions 
proportional to the ratio $a_0/a_1$.
For example, the coefficient of $L^{FF}_1$ in (\ref{logdivC2}),
entering the beta function of $c^{FF}_1$, would become
$$
\frac{71}{15}+\frac58\frac{a_0}{a_1}+\frac{7}{48}\frac{a_0^2}{a_1^2}\ .
$$
This will not change the general structure of the divergences.
The lower order equation of motion of the connection will still
imply the vanishing of torsion,
but the lower order equation of motion of the metric will now
additionally imply that the Ricci tensor must be zero.
Then the possible counterterms will be the same as in GR,
except that now they will appear already at one loop.

The discussions in \cite{Tseytlin:1981nu} offer some additional hints.
For example, it was noted that the theory with Lagrangian
\be
\frac12 m_P^2 R
+c^{FF}_1  F_{\alpha\beta\gamma\delta}F^{\alpha\beta\gamma\delta}
\ee
is renormalizable when one sets $T=0$, since then it reduces to QG,
but not in general, where it amounts to GR coupled to a Yang-Mills field,
a system that was studied in \cite{Deser:1974xq}.
It seems plausible that the theory will have better chances of being
renormalizable when its Einstein form will contain the QG operators.
In this connection it is worth noting that the known examples of 
ghost- and tachyon-free MAGs
do not contain the QG operators in their Einstein form.

Renormalizability demands that the kinetic term of $Z$ contain two derivatives
but that of $X$ contain four derivatives
\cite{sezgin,Lin:2018awc,Lin:2019ugq}.
This can be achieved by adding to the action terms of the form $(DT)^2$,
in the tetrad formalism.
For example, the Lagrangian 
\be
c^{FF}_1 F_{\alpha\beta\gamma\delta}F^{\alpha\beta\gamma\delta}
+c^{TT}_1  D_\alpha T_{\beta\gamma\delta}D^\alpha T^{\beta\gamma\delta}
\ee
has a kinetic operator that contains in its leading $X-X$ part an additional power
of $\nabla^2$ compared to the theory we studied in this paper.
The elimination of nonminimal terms by our gauge choice
would still be possible for this kinetic operator.
How generally it can be made to work is an open question.

Besides this very technical point, as already noticed in \cite{Baldazzi:2021kaf}, 
the number of derivatives depends on the gauge
and so could the power counting.
This would be analogous to what happens in spontaneously broken gauge theories,
where renormalizability is proven in one gauge and unitarity in another.
Indeed we know that working with coordinate frames or orthonormal frames
amounts to a choice of unitary gauge in a more general
$GL(4)$-invariant formulation of the theory,
in the sense that a field transforming under $GL(4)$ by a shift is set to zero.
Thus, it seems likely that among all MAGs there will be some that are
renormalizable, but it may be necessary to choose a special gauge
to see this.

\begin{appendix}

\section{Partial results}

\subsection{The contribution of $X$ and $Z$}

The commutator of the Riemannian covariant derivatives on the fields are
\be
[\nabla_\mu,\nabla_\nu]\begin{pmatrix}
X\\Z
\end{pmatrix}
=\begin{pmatrix}
\Omega_{XX\mu\nu} &0\\
0 &\Omega_{ZZ\mu\nu}
\end{pmatrix}
\begin{pmatrix}
X\\Z
\end{pmatrix}
\ee
with
\bea
\Omega_{XX\mu\nu} {}^{\alpha\beta}{}_{\rho\sigma}
&=&R_{\mu\nu}{}^\alpha{}_\rho \delta^\beta_\sigma
+R_{\mu\nu}{}^\beta{}_\sigma \delta^\alpha_\rho
\\
\Omega_{ZZ\mu\nu} {}^{\alpha\beta\gamma}{}_{\rho\sigma\lambda}
&=&R_{\mu\nu}{}^\alpha{}_\rho \delta^\beta_\sigma \delta^\gamma_\lambda
+R_{\mu\nu}{}^\beta{}_\sigma \delta^\alpha_\rho \delta^\gamma_\lambda
+R_{\mu\nu}{}^\gamma{}_\lambda \delta^\alpha_\rho \delta^\beta_\sigma
\eea

Then one can calculate the terms entering in (\ref{logdiv}).
We have the following dimension-two parts
\bea
\tr W&=&10R-3\nabla_\mu T^\mu+\ldots
\nonumber\\
\tr\nabla_\mu V^\mu&=&0
\nonumber\\
\tr V_\mu V^\mu&=&-\tfrac{1}{12}F_{\mu\nu}{}^{\mu\nu}
+\ldots\ ,
\nonumber
\eea
where the dots stand for terms that do not contribute
to the terms we are interested in, and
\bea
\tr W^2&=&
9 H^{RR}_1
+10 H^{RR}_2
+34H^{RT}_1
+15H^{RT}_2
+4H^{RT}_3
+4H^{RT}_4
+2H^{RT}_5
\nonumber\\
&&
+\tfrac{25}{2} H^{TT}_1
+7 H^{TT}_2
-2 H^{TT}_3
-\tfrac{23}{2} H^{TT}_4
+\tfrac{29}{2} H^{TT}_5
-\tfrac{29}{4} H^{TT}_6
\nonumber\\
&&
-24 H^{TT}_7
+8 H^{TT}_8
+2 H^{TT}_9+\ldots
\nonumber\\
\tr W\bar\nabla_\mu V^\mu&=&
-\tfrac{13}{4}H^{TT}_1
-\tfrac32H^{TT}_2
-\tfrac{17}{2} H^{TT}_4
\nonumber\\
&&
+\tfrac{11}{2} H^{TT}_5
-\tfrac{19}{4} H^{TT}_6
-10 H^{TT}_7
+4 H^{TT}_8
+2 H^{TT}_9+\ldots
\nonumber\\
\tr \nabla_\mu V^\mu \nabla_\nu V^\nu&=&
-\tfrac{13}{2}H^{TT}_1
-3H^{TT}_2
-17 H^{TT}_4
\nonumber\\
&&
+11 H^{TT}_5
-\tfrac{19}{2} H^{TT}_6
-20 H^{TT}_7
+8 H^{TT}_8
+4 H^{TT}_9+\ldots
\nonumber\\
\tr\bar\nabla_\mu V_\nu\bar\nabla^\mu V^\nu
&=&
-\tfrac{105}{2} H^{TT}_1
-21 H^{TT}_2
+12 H^{TT}_3+\ldots
\nonumber\\
\tr\bar\nabla_\mu V_\nu\bar\nabla^\nu V^\mu
&=&
-\tfrac{13}{2} H^{TT}_1
-3 H^{TT}_2
-17 H^{TT}_4
+11 H^{TT}_5
-\tfrac{19}{2} H^{TT}_6
-20 H^{TT}_7
+8 H^{TT}_8
+4 H^{TT}_9+\ldots
\nonumber\\
\tr\Omega_{\mu\nu}\Omega^{\mu\nu}
&=&-22 H^{RR}_1+\ldots
\nonumber\\
\tr\Omega_{\mu\nu}\nabla^\mu V^\nu
&=&8H^{RT}_1+11 H^{RT}_2-2 H^{RT}_4+\ldots
\nonumber\\
\tr\Omega_{\mu\nu}V^\mu V^\nu
&=&6H^{RR}_2+6H^{RT}_3-6H^{RT}_4+\ldots
\nonumber\\
\tr W V^\mu V_\mu
&=&
-\tfrac92 H^{RR}_1
-10 H^{RR}_2
-4 H^{RR}_3
\nonumber\\
&&
-67H^{RT}_1
-\tfrac{17}{2}H^{RT}_2
-18H^{RT}_3
+6H^{RT}_4
-10H^{RT}_5
\nonumber\\
&&
-\tfrac{53}{2} H^{TT}_1
-13 H^{TT}_2
+2H^{TT}_3
+16H^{TT}_4
-24 H^{TT}_5
\nonumber\\
&&
+\tfrac{29}{4}H^{TT}_6
+\tfrac{63}{2}H^{TT}_7
+2H^{TT}_8
-2H^{TT}_9+\ldots
\nonumber\\
\tr\nabla_\mu V^\mu V_\nu V^\nu
&=&
0+\ldots
\nonumber\\
\tr\nabla_\mu V_\nu [V^\mu, V^\nu]
&=&
-54 H^{RT}_1
+39 H^{RT}_2
-36 H^{RT}_3
+16H^{RT}_4
+4H^{RT}_5
\nonumber\\
&&
-\tfrac{51}{2} H^{TT}_1
-9H^{TT}_2
+12H^{TT}_3
+45H^{TT}_4
+3 H^{TT}_5
\nonumber\\
&&
+13H^{TT}_6
+H^{TT}_7
-24H^{TT}_8
-20H^{TT}_9+\ldots
\nonumber\\
\tr V_\mu V^\mu V_\rho V^\rho 
&=&
99H^{RR}_1
-6H^{RR}_2
+4H^{RR}_3
\nonumber\\
&&
+412H^{RT}_1
+190H^{RT}_2
-12H^{RT}_3
+12H^{RT}_4
+16H^{RT}_5
\nonumber\\
&&
+\tfrac{287}{2} H^{TT}_1
+93 H^{TT}_2
-16H^{TT}_3
-92H^{TT}_4
+102 H^{TT}_5
\nonumber\\
&&
-59 H^{TT}_6
-186H^{TT}_7
+32 H^{TT}_8
+26 H^{TT}_9+\ldots
\nonumber\\
\tr\,[V_\mu,V_\nu][V^\mu,V^\nu]
&=&
-222 H^{RR}_1
+108 H^{RR}_2
-8H^{RR}_3
\nonumber\\
&&
-1064H^{RT}_1
-356H^{RT}_2
+216H^{RT}_3
-216H^{RT}_4
-32H^{RT}_5
\nonumber\\
&&
-359 H^{TT}_1
-210 H^{TT}_2
+128H^{TT}_3
+328 H^{TT}_4
-180 H^{TT}_5
\nonumber\\
&&
+154 H^{TT}_6
+420 H^{TT}_7
-256 H^{TT}_8
-52 H^{TT}_9+\ldots
\eea
Inserting these in (\ref{logdiv}) we find
\bea
B_{XZ}&=&
-\frac12\frac{1}{(4\pi)^2}\int d^4x\sqrt{g}\Big[
\frac{533}{144}H^{RR}_1
+\frac{227}{72}H^{RR}_2
-\frac{55}{36}H^{RR}_3
\nonumber\\
&&\qquad\qquad\qquad\qquad
+\frac{17}{2}H^{RT}_1
+6H^{RT}_2
+\frac14 H^{RT}_3
+\frac{23}{12}H^{RT}_4
-\frac{5}{6}H^{RT}_5
\nonumber\\
&&\qquad\qquad\qquad\qquad
+\frac{211}{96}H^{TT}_1
+\frac{33}{16}H^{TT}_2
-\frac13H^{TT}_3
-\frac{47}{24}H^{TT}_4
+\frac{37}{24}H^{TT}_5
\nonumber\\
&&\qquad\qquad\qquad\qquad
-\frac{29}{16}H^{TT}_6
-\frac{107}{24}H^{TT}_7
+\frac{23}{6} H^{TT}_8
+\frac{29}{24}H^{TT}_9
+\ldots\Big]\ .
\label{XZtotal}
\eea

%%%%%%%%%%%%%%%%%%%%
\subsection{The contribution of the ghosts}
%%%%%%%%%%%%%%%%%%%%%

The ghost operator can be written in the form
\be
-\begin{pmatrix}
1& 0\\
0 & 1
\end{pmatrix}
\nabla^2
+
\begin{pmatrix}
V_{LL}^\mu& V_{LD}^\mu\\
V_{DL}^\mu & V_{DD}^\mu
\end{pmatrix}
\nabla_\mu
+
\begin{pmatrix}
W_{LL}^\mu& W_{LD}^\mu\\
W_{DL}^\mu & W_{DD}^\mu
\end{pmatrix}\  ,
\ee
where the coefficients $V$ and $W$ can be read off from (\ref{fpgh}).
We observe the special property
$$
\nabla_\mu V_{AB}^\mu=W_{AB}\ ,
$$
for $A,B=L,D$.

We also we need the commutator of the Riemannian covariant derivatives
\be
[\nabla_\mu,\nabla_\nu]\begin{pmatrix}
\Sigma\\C
\end{pmatrix}
=\begin{pmatrix}
\Omega_{LL\mu\nu} &0\\
0 &\Omega_{DD\mu\nu}
\end{pmatrix}
\begin{pmatrix}
\Sigma\\C
\end{pmatrix}
\ee
with
\bea
\Omega_{LL\mu\nu} {}^{\alpha\beta}{}_{\rho\sigma}
&=&2R_{\mu\nu}{}^{[\alpha}{}_{[\rho}\delta^{\beta]}_{\sigma]}
\\
\Omega_{DD\mu\nu} {}^{\alpha}{}_{\beta}
&=&R_{\mu\nu}{}^\alpha{}_\beta\ .
\eea

Then one can calculate the terms entering in (\ref{logdiv}):
\bea
\tr W^2&=&
-\frac12 H^{TT}_4+\frac32 H^{TT}_5-\frac34 H^{TT}_6-2 H^{TT}_7+\ldots
\nonumber\\
\tr W\bar\nabla_\mu V^\mu&=&
-\frac12 H^{TT}_4+\frac32 H^{TT}_5-\frac34 H^{TT}_6-2 H^{TT}_7+\ldots
\nonumber\\
\tr \nabla_\mu V^\mu \nabla_\nu V^\nu&=&
-\frac12 H^{TT}_4+\frac32 H^{TT}_5-\frac34 H^{TT}_6-2 H^{TT}_7+\ldots
\nonumber\\
\tr\bar\nabla_\mu V_\nu\bar\nabla^\mu V^\nu
&=&
-\frac54 H^{TT}_1-\frac12 H^{TT}_2+\ldots
\nonumber\\
\tr\bar\nabla_\mu V_\nu\bar\nabla^\nu V^\mu
&=&
-\frac12 H^{TT}_4+\frac32 H^{TT}_5-\frac34 H^{TT}_6-2 H^{TT}_7+\ldots
\nonumber\\
\tr\Omega_{\mu\nu}\Omega^{\mu\nu}
&=&-3 H^{RR}_1+\ldots
\nonumber\\
\tr\Omega_{\mu\nu}\nabla^\mu V^\nu
&=&2H^{RT}_1+\frac32H^{RT}_2+\ldots
\nonumber\\
\tr\Omega_{\mu\nu}V^\mu V^\nu
&=&H^{RR}_2+H^{RT}_3-H^{RT}_4+\ldots
\nonumber\\
\tr W V^\mu V_\mu
&=&
-H^{RT}_3
-H^{RT}_4
+H^{TT}_3
-\frac12 H^{TT}_4
-\frac12 H^{TT}_5
+\frac34 H^{TT}_6
+H^{TT}_7
-3H^{TT}_8+\ldots
\nonumber\\
\tr \nabla_\mu V^\mu V_\nu V^\nu&=&
-H^{RT}_3
-H^{RT}_4
+H^{TT}_3
-\frac12 H^{TT}_4
-\frac12 H^{TT}_5
+\frac34 H^{TT}_6
+H^{TT}_7
-3H^{TT}_8+\ldots
\nonumber\\
\tr\nabla_\mu V_\nu [V^\mu, V^\nu]
&=&
-2H^{RT}_1
+H^{RT}_2
-H^{RT}_3
+H^{RT}_4
\nonumber\\
&&
+\frac12 H^{TT}_1
+H^{TT}_2
-H^{TT}_3-\frac32 H^{TT}_4-\frac32 H^{TT}_5
+\frac14 H^{TT}_6-H^{TT}_7+H^{TT}_8+\ldots
\nonumber\\
\tr V_\mu V^\mu V_\rho V^\rho 
&=&%F_{abcd}F^{cdab}+F^{(13)}_{ab}F^{(13)ba}
H^{RR}_1+H^{RR}_2
+4H^{RT}_1+2H^{RT}_2
+2H^{RT}_3-2H^{RT}_4
\nonumber\\
&&
+\frac52 H^{TT}_4+\frac52 H^{TT}_5
-\frac54 H^{TT}_6+H^{TT}_9+\ldots
\nonumber\\
\tr\,[V_\mu,V_\nu][V^\mu,V^\nu]
&=&%2F_{abcd}F^{cdab}+4F_{abcd}F^{acbd}-2F^{(13)}_{ab}F^{(13)ba}
-2H^{RR}_2
-8H^{RT}_1+4H^{RT}_2
-4H^{RT}_3+4H^{RT}_4
\nonumber\\
&&
+H^{TT}_1+2 H^{TT}_2
-3 H^{TT}_4-3 H^{TT}_5
+\frac32 H^{TT}_6-4H^{TT}_7-2H^{TT}_9+\ldots
\eea
where once again we omit terms containing higher powers of $R$, $T$.
Summing up everything we get
\bea
B_{gh}&=&
-\frac12\frac{1}{(4\pi)^2}\int d^4x\sqrt{g}\Big[
-\frac{203}{1440}H^{RR}_1
+\frac{19}{720}H^{RR}_2
+\frac{5}{18}H^{RR}_3
\nonumber\\
&&\qquad\qquad\qquad\qquad
-\frac{1}{6}H^{RT}_1
-\frac{5}{24}H^{RT}_2
+\frac{1}{24}H^{RT}_3
-\frac{7}{24}H^{RT}_4
+\frac{1}{12}H^{RT}_5
\nonumber\\
&&\qquad\qquad\qquad\qquad
-\frac{13}{192}H^{TT}_1
-\frac{5}{96}H^{TT}_2
+\frac{1}{6}H^{TT}_3
+\frac{1}{48}H^{TT}_4
+\frac{3}{16}H^{TT}_5
\nonumber\\
&&\qquad\qquad\qquad\qquad
-\frac{1}{96}H^{TT}_6
-\frac{1}{48}H^{TT}_7
-\frac{5}{12} H^{TT}_8
+\frac{1}{48}H^{TT}_9+\ldots\Big]\ .
\label{ghosttotal}
\eea

\end{appendix}

\goodbreak

\end{document}